# Heat transport as a probe of superconducting gap structure

H Shakeripour[1], C Petrovic[2,3] and Louis Taillefer[1,3]

[1] Département de physique and RQMP, Université de Sherbrooke, Sherbrooke, Québec J1K 2R1, Canada

[2] Condensed Matter Physics and Materials Science Department, Brookhaven National Laboratory, Upton, New York 11973, USA

[3] Canadian Institute for Advanced Research, Toronto, Ontario M5G 1Z8, Canada

E-mail: Louis.Taillefer@USherbrooke.ca

**Abstract**. The structure of the superconducting gap provides important clues on the symmetry of the order parameter and the pairing mechanism. The presence of nodes in the gap function imposed by symmetry implies an unconventional order parameter, other than *s*-wave. Here we show how measurements of the thermal conductivity at very low temperature can be used to determine whether such nodes are present in a particular superconductor, and shed light on their nature and location. We focus on the residual linear term at $T \rightarrow 0$. A finite value in zero magnetic field is strong evidence for symmetry-imposed nodes, and the dependence on impurity scattering can distinguish between a line of nodes or point nodes. Application of a magnetic field probes the low-energy quasiparticle excitations, whether associated with nodes or with a small value of the gap on some part of the Fermi surface, as in a multi-band superconductor. We frame our discussion around archetypal materials: Nb for *s*-wave, $Tl_2Ba_2CuO_{6+\delta}$ for *d*-wave, $Sr_2RuO_4$ for *p*-wave, and $NbSe_2$ for multi-band superconductivity. In that framework, we discuss three heavy-fermion superconductors: $CeIrIn_5$, $CeCoIn_5$ and $UPt_3$.

## 1. Residual heat conduction as $T \rightarrow 0$

A superconductor is a perfect conductor of charge, but a poor conductor of heat. Indeed, in the limit of zero temperature, electronic heat conduction in a fully gapped superconductor goes to zero, as there are no thermally-excited quasiparticles to carry heat (the pairs in the condensate carry no heat). This shows up as a thermal conductivity $\kappa(T)$ whose linear term $\kappa / T$ goes to zero as the temperature $T$ goes to zero, *i.e.* $\kappa / T \rightarrow 0$ as $T \rightarrow 0$, or $\kappa_0 / T = 0$. An example of this is shown in Figure 1, where the thermal conductivity of $NbSe_2$ is plotted as $\kappa / T$ vs $T^2$. The data extrapolates linearly to zero at $T = 0$ [1]. This confirms $NbSe_2$ as a fully-gapped superconductor, in agreement with the general consensus on an order parameter with *s*-wave symmetry in this material [2, 3].

The situation is different if the gap function goes to zero for certain directions imposed by symmetry (*i.e.* with an associated sign change in the wavefunction). Impurity scattering broadens those nodes to produce a finite residual density of states at zero energy. The associated low-energy quasiparticle excitations can conduct heat, and hence $\kappa_0 / T > 0$ [4]. Note that if the nodes in the gap are accidental, not imposed by symmetry (as in the so-called "extended *s*-wave gap"), impurity scattering will lift rather than broaden the nodes and produce a (more isotropic) finite gap [5]. The consequence on heat conduction is thus qualitatively different, as $\kappa_0 / T = 0$ in this case.

The clearest case is that of high-$T_c$ superconductors, for which the order parameter symmetry is well-established to be *d*-wave. In the strongly overdoped regime, the simplest of these quasi-2D materials have a Fermi surface that consists of a single large hole-like cylinder. The *d*-wave symmetry requires that the gap vanish along four vertical lines running along the axis of the cylinder. Line nodes in the gap produce a density of states that grows linearly with energy at low energy. This leads to universal heat conduction at $T \rightarrow 0$, a remarkable property whereby the increase in the zero-energy density of states caused by impurity scattering is exactly compensated by a corresponding decrease in mean free path, making the thermal conductivity independent of impurity scattering (in the clean limit) [4, 6]. Universal heat conduction was first verified in the high-$T_c$ superconductors $YBa_2Cu_3O_y$ (YBCO) [7] and $Bi_2Sr_2CaCu_2O_{8+\delta}$ (Bi-2212) [8]. In Figure 1, we show data for strongly overdoped $Tl_2Ba_2CuO_{6+\delta}$ (Tl-2201), with $T_c = 15$ K [9]. The residual linear term is unambiguous.

Universal heat conduction is thus seen to be a powerful probe of order-parameter symmetry, for its observation in a material is compelling evidence that the superconducting gap has a line of nodes. The only known exception to this rule is the quadratic point nodes that can occur along certain directions in special symmetries [10] (*e.g.* along the *c*-axis in the (1, *i*) state of the odd-parity $E_{2u}$ representation in $D_{6h}$ hexagonal symmetry [4].) Theoretically, universal heat conduction was shown to be unaffected by vertex and Fermi-liquid corrections [6]. It is given simply by [4]:

$$\kappa_0 / T = ( \pi^2 k^2_B / 3 ) \ N_F v^2_F \ ( a \hbar / 2 \mu \Delta_0 ) \qquad (1)$$

where $N_F$ is the density of states at the Fermi energy, $v_F$ the Fermi velocity at the node, $\mu\Delta_0$ the slope of the gap at the node (*e.g.* for a *d*-wave gap in 2D, $\Delta = \Delta_0 \cos2\varphi$, $\mu = 2$ [4]), and *a* is a parameter of order unity ($a = 4 / \pi$ for *d*-wave in 2D; $a = 1$ for a line in the basal plane in 3D [4]). Applying Eq. (1) to Tl-2201 reveals good quantitative agreement between theory and experiment for the magnitude of $\kappa_0 / T$ [11]. This means that from a knowledge of the gap magnitude (*e.g.* from $T_c$) and the Fermi surface, one can predict, using Eq. (1), the magnitude of $\kappa_0 / T$ to be expected if the superconducting gap has a line node. This was done for the quasi-2D oxide superconductor $Sr_2RuO_4$ [12], believed to have *p*-wave symmetry [13], for which $N_F$ and $v_F$ are known precisely [13]. The thermal conductivity of $Sr_2RuO_4$, reproduced from [12] in Figure 1, was shown to indeed be universal, and the magnitude of $\kappa_0 / T$ in the clean limit is in excellent agreement with theoretical expectation [12].

If the gap vanishes at point nodes rather than line nodes, conduction is not universal (except in the special case of quadratic point nodes mentioned above), and $\kappa_0 / T$ grows with impurity scattering [4, 10]. Although in very clean samples (where the impurity scattering rate $\Gamma \rightarrow 0$) $\kappa_0 / T \rightarrow 0$, in any real sample $\kappa_0 / T > 0$. In other words, quite generally, the observation of a non-zero $\kappa_0 / T$ is strong evidence for the presence of nodes in the gap. There is one caveat: the sample must be homogeneous, with no non-superconducting regions, as these would give a fraction of the normal-state residual linear term. Conversely, the absence of a residual linear term, *i.e.* $\kappa_0 / T \approx 0$, is strong evidence that the gap has no symmetry-imposed nodes.

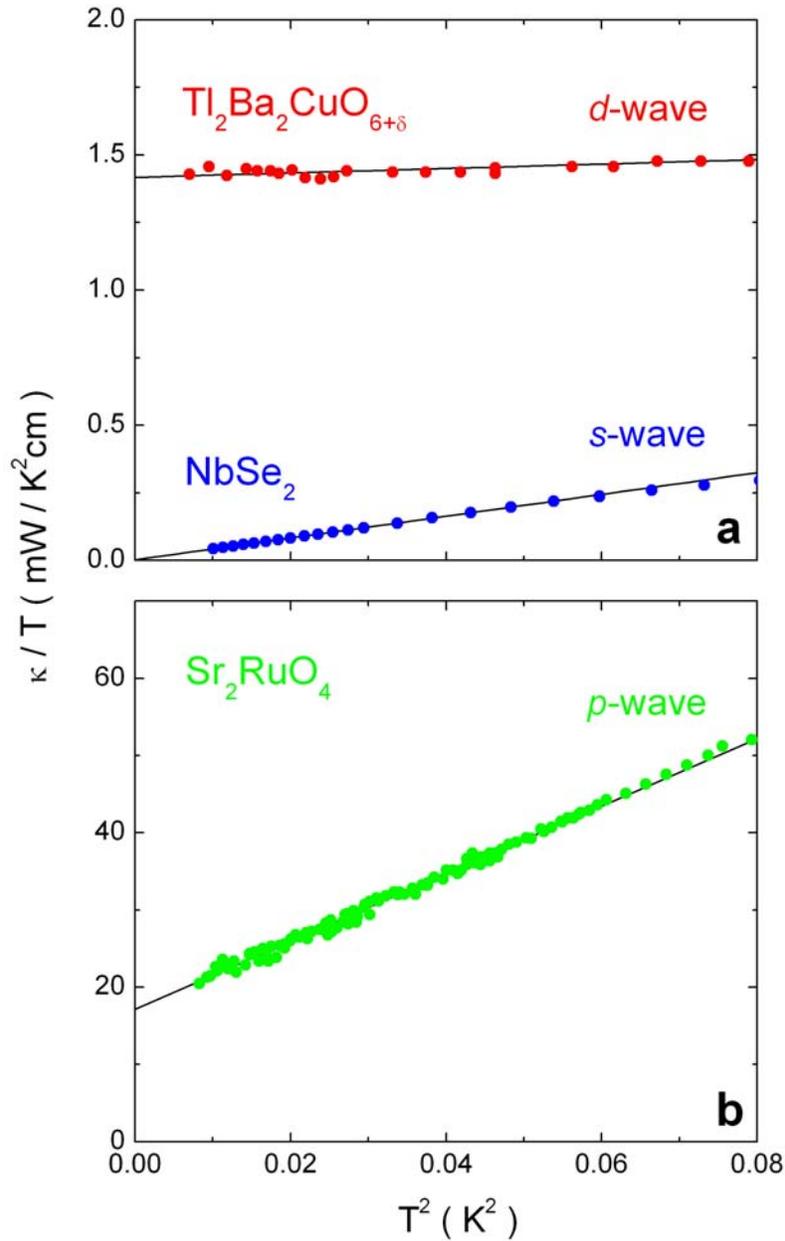

**Figure 1. Thermal conductivity of archetypal superconductors.**

**a)** Thermal conductivity of quasi-2D multiband *s*-wave superconductor NbSe$_2$ [1] and quasi-2D single-band *d*-wave superconductor Tl-2201 [9, 11]. Note the large residual linear term $\kappa / T$ as $T \to 0$ in the *d*-wave superconductor, due to nodal quasiparticle excitations. **b)** Thermal conductivity in the quasi-2D three-band superconductor Sr$_2$RuO$_4$ [12], believed to have *p*-wave symmetry.

## 2. Magnetic field dependence

Application of a magnetic field $H$ provides complementary information on the gap structure. Here again, the most diagnostic regime is the limit of $T \rightarrow 0$. The field is a way to excite quasiparticles in a type-II superconductor, even at $T = 0$, and probe the excitation spectrum at finite energy. In an $s$-wave superconductor at $T = 0$, the quasiparticles are localized in states within the vortex core and any conduction perpendicular to the magnetic field is due to tunnelling between adjacent vortices. This tunnelling process leads to a slow exponential growth of $\kappa_0 / T$ vs $H$, as shown for Nb in Figure 2. This slow rise of $\kappa_0 / T$ as a function of field at $H \ll H_{c2}(0)$ is a typical signature of an $s$-wave superconductor (or more precisely a superconductor with a nodeless gap). Provided the $s$-wave gap does not have strong band or angle dependence, this will cause $\kappa_0 / T$ vs $H$ to have an upward (concave) curvature (see Figure 2).

This textbook field dependence can be strongly modified if the $s$-wave gap does have pronounced band and / or angle dependence. The classic examples of an $s$-wave superconductor whose gap is very different on two parts of the Fermi surface are $MgB_2$ and $NbSe_2$. In these so-called "multi-band superconductors", it is effectively as though each Fermi surface has its own $H_{c2}$. In $NbSe_2$, the gap is estimated to differ by a factor ~ 3 between the Nb $d$-band Fermi surface (where it is large) and the Se $p$-band Fermi surface (where it is small) [2, 3]. This means that a field $H^* \approx H_{c2}(0) / 9$ should be sufficient to suppress superconductivity on the weak-gap Fermi surface. This shows up as a shoulder in the $\kappa_0 / T$ vs $H$ curve at $H^*$, seen in both $MgB_2$ [14] and $NbSe_2$ [1] (see Figure 2). By zooming on the very low field region where $H \ll H^*$, one can still see the upward (concave) curvature typical of $s$-wave superconductors [1], but the overall dependence up to $H_{c2}(0)$ is very different from the textbook dependence (see Figure 2).

In a nodal superconductor, there are delocalized states outside the vortex cores even at $T = 0$ and these dominate the transport in the vortex state. In this case, the effect of a magnetic field is to shift the low-energy states and thus excite more quasiparticles. There are two mechanisms for this shift: 1) the Doppler shift due to superfluid flow around the vortices, which yields a $H^{1/2}$ growth in quasiparticle density of states (the Volovik effect [15]); 2) the shift due to a coupling to spin, which yields a linear growth in the density of states (the Zeeman effect [16]). At low fields, the former dominates, and it causes $\kappa_0 / T$ to rise rapidly. The textbook case is the 2D $d$-wave superconductor, illustrated in Figure 2 with data on strongly overdoped Tl-2201 (from [9]).

## 3. Some examples

We can now summarize the two basic characteristics of heat transport in archetypal $s$-wave and $d$-wave superconductors, displayed in Figure 2: in the former, $\kappa_0 / T = 0$ at $H = 0$ and grows slowly with $H$ for $H \ll H_{c2}(0)$; in the latter, $\kappa_0 / T > 0$ at $H = 0$ and grows rapidly with $H$ for $H \ll H_{c2}(0)$. With the added caveat that in a superconductor with multiple Fermi surface sheets, the low-field regime might be limited to much lower fields, *i.e.* $H \lll H_{c2}(0)$.

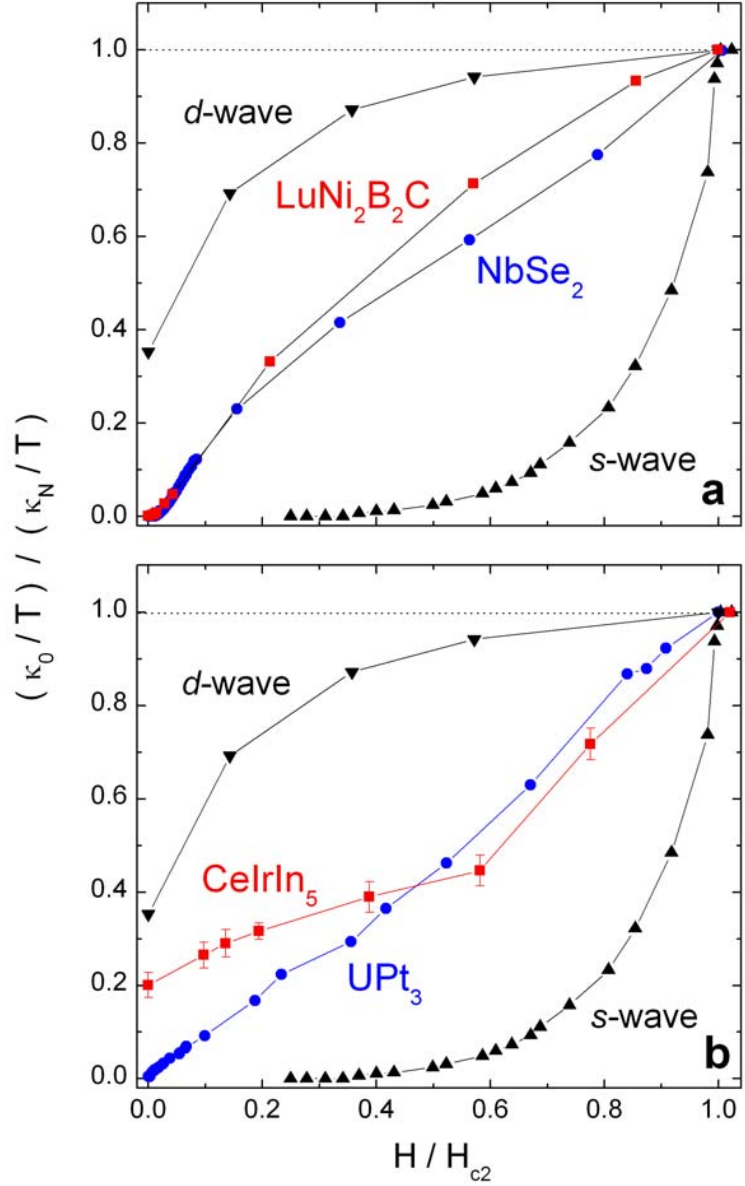

**Figure 2. Field dependence of thermal conductivity in the $T = 0$ limit.**

Residual linear term $\kappa_0 / T$ in the thermal conductivity vs magnetic field $H$, normalized at $H_{c2}$. The textbook behaviour for $s$-wave and $d$-wave superconductors is shown in black symbols: for elemental Nb (up triangles; [34]) and the single-band overdoped cuprate Tl-2201 (down triangles; [9]), respectively. **a)** Quasi-2D multi-band $s$-wave superconductor NbSe$_2$ [1] and multi-band borocarbide LuNi$_2$B$_2$C [19]. **b)** Two heavy-fermion superconductors: tetragonal CeIrIn$_5$ (this work) and hexagonal UPt$_3$ [33]. In all cases, $\mathbf{J} \perp \mathbf{H}$ and $\mathbf{H} \parallel \mathbf{c}$ (except for UPt$_3$ where $\mathbf{J} \parallel \mathbf{H} \parallel \mathbf{a}$).

By looking for these signatures in a newly discovered superconductor, one can gain insight into its gap structure with a fairly simple bulk measurement. For instance, this was done for two classes of superconductors discovered recently: the intercalated graphite compound $C_6Yb$ [17] and Cu-doped $TiSe_2$ [18]. In both cases, the two classic signatures of an *s*-wave gap were observed: 1) $\kappa_0 / T = 0$ at $H = 0$; 2) a slow growth at low $H$ with upward (concave) curvature. In neither case was there evidence of strong band (or angle) dependence. Because most superconductors will in general have several Fermi surface sheets (and a non-cubic crystal structure), one should always keep in mind the possibility that the gap amplitude will vary from sheet to sheet (or with angle). The borocarbide superconductors are an example of this. In Figure 2, we reproduce data for $LuNi_2B_2C$ [19], which looks remarkably similar to $NbSe_2$: $\kappa_0 / T = 0$ at $H = 0$, but then grows rapidly with $H$. There are two possible interpretations: 1) *s*-wave with a very small minimum gap (roughly 10 times smaller than the gap maximum [19]) somewhere on the Fermi surface (possibly from strong angle dependence); 2) unconventional symmetry (non *s*-wave), with point nodes. The second scenario is a possibility only if the $LuNi_2B_2C$ crystal investigated in [19] was in the very clean limit. A test of this scenario would therefore be to measure the thermal conductivity of $LuNi_2B_2C$ in samples with stronger impurity scattering (larger residual resistivity). Barring this special ultraclean limit, the first scenario seems the most likely, especially as the superconducting gap of $YNi_2B_2C$ was recently shown directly by angle-resolved photoemission spectroscopy to be strongly band and angle dependent [20].

## 4. Heavy-fermion superconductors

Let us now discuss three of the most studied heavy-fermion superconductors: $CeIrIn_5$ (tetragonal, $T_c = 0.4$ K); $CeCoIn_5$ (tetragonal, $T_c = 2.4$ K); $UPt_3$ (hexagonal, $T_c = 0.44$ K). The former two are part of a family of materials that also includes $CeRhIn_5$, an antiferromagnet which becomes a superconductor under pressure. As is the case for most heavy-fermion superconductors, these three superconductors are all close to an antiferromagnetic instability [21, 22]. This strongly suggests that pairing in heavy-fermion superconductors is magnetically mediated [23, 24]. However, progress in understanding the pairing mechanism has been hampered by the fact that the symmetry of the order parameter is not yet definitively established in any heavy-fermion superconductor.

*4.1. CeIrIn$_5$*

In this context, the technique of heat transport can play a powerful role, as it is one of very few *directional* probes, sensitive not only to the presence of nodes in the gap but also to their location around the Fermi surface. This directional sensitivity is illustrated in Figure 3, where we reproduce data for $CeIrIn_5$ [25, 26]. Conduction is qualitatively different for heat flow parallel and perpendicular to the high-symmetry (tetragonal) *c*-axis. For a current in the basal plane (***J*** ∥ ***a***), there is a large residual linear term $\kappa_0 / T$ [25], independent of impurity scattering [26]. This universal heat conductivity is in quantitative agreement with theoretical expectation for a line node [25, 26]. For a current normal to the basal plane (***J*** ∥ ***c***), the residual linear term is negligible in pure crystals [25], and $\kappa_0 / T$ grows with impurity scattering [26]. All this constitutes compelling evidence for a line node in the basal plane. The presence or not of a point node along the *c*-axis is more difficult to ascertain, and gaps with and without such point nodes, such as the "hybrid" and "polar" gaps illustrated in Figure 3, are both consistent with the data. A "horizontal" line node like this is not easily compatible with the *d*-wave symmetry proposed for both $CeIrIn_5$ and $CeCoIn_5$, as in such a symmetry the line nodes would be "vertical", located where the Fermi surface intersects "vertical" planes such as the *ac* plane (see [26] and references therein).

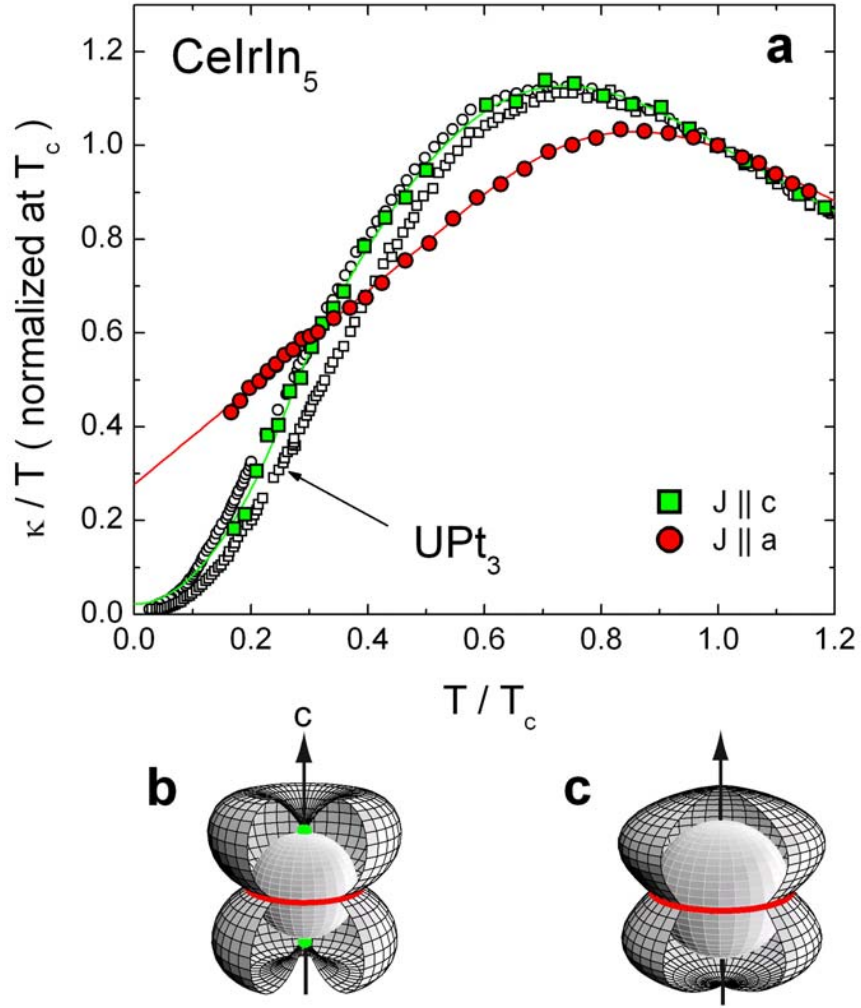

**Figure 3. Anisotropy of heat conduction in the heavy-fermion superconductors CeIrIn$_5$ and UPt$_3$.**

**a)** In-plane (circles) and out-of-plane (squares) thermal conductivity of tetragonal CeIrIn$_5$ [25, 26] and hexagonal UPt$_3$ [33], normalized at $T_c$. The data for $c$-axis transport in CeIrIn$_5$ tracks the data of both $b$- and $c$-axis transport in UPt$_3$, while the $a$-axis data in CeIrIn$_5$ is qualitatively different. Lines are guides to the eye. **b)** Schematic of the "hybrid gap", $\Delta^2 \sim \cos^2\theta \sin^2\theta$ [10], with a horizontal line node in the $ab$-plane and two linear point nodes along the $c$-axis (black arrow), drawn on a single spherical Fermi surface. **c)** Schematic of the "polar gap", $\Delta^2 \sim \cos^2\theta$ [10], which only has a line node in the $ab$-plane. $\theta$ is the polar angle away from the $c$-axis.

In Figure 2, we present data on the field dependence of $\kappa_0 / T$ in CeIrIn$_5$. Clearly non-monotonic, it breaks up into two regimes, below and above $H^* \approx 0.6\ H_{c2}$. Below $H^*$, $\kappa_0 / T$ rises roughly linearly, with perhaps a slight downward (convex) curvature. But an extrapolation of this dependence up to $H_{c2}$ would only give half the normal-state conductivity. The remaining half of the increase occurs above $H^*$. This suggests a multi-band scenario whereby the gap magnitude may be quite different on different parts of the Fermi surface. Note that the complex Fermi surface of CeIrIn$_5$ consists of 4 or 5 sheets, with varying degree of $f$-character and $c$-axis dispersion [27, 28].

*4.2. CeCoIn$_5$*

Let us turn to the isostructural superconductor CeCoIn$_5$. Heat transport studies as a function of impurity doping showed this material to be an extreme case of multi-band superconductivity, whereby part of the Fermi surface has a negligible gap [29]. This was confirmed by further studies [30], which showed superconductivity on the small-gap Fermi surface sheet to be destroyed by application of a minute magnetic field ($H \sim H_{c2}/1000$). This sheet therefore shows metallic behaviour even deep in the superconducting state. In this context, it is difficult to ascertain the structure of the large gap from a measurement of the thermal conductivity. The coexistence of antiferromagnetic order with superconductivity in CeCoIn$_5$ [31] may be responsible for this extreme band dependence of the superconducting gap.

*4.3. UPt$_3$*

The superconductor UPt$_3$ is an interesting, unresolved case. Besides the well-established fact that its field-temperature phase diagram consists of three separate superconducting phases [22], somewhat reminiscent of superfluid helium-3, the symmetry of the order parameter in any of these phases is undetermined. While there is overwhelming evidence for nodes in the gap [22], the nature and location of these nodes are still unclear. Several experiments have been interpreted in terms of a line node in the basal plane, but, as we shall see, measurements of heat transport [32] rule this out.

In Figure 3, we compare our data on pure CeIrIn$_5$ with the data of [33] on pure UPt$_3$. The two compounds are very similar: not only is the value of $T_c$ the same, but the normal-state heat transport is essentially identical, with the same residual resistivity in the best crystals ($\rho_0 \approx 0.2\ \mu\Omega$ cm), the same strength of inelastic scattering, and the same temperature-independent anisotropy, albeit in reverse order ($\kappa_{Nc}/\kappa_{Nb} = 2.7$). The electronic specific heat in UPt$_3$ is a factor 1.4 larger ($\gamma_N = 1.04 \times 10^4$ J / K$^2$ mole). However, the in-plane Fermi velocity $v_F$ is quite different: $v_{Fa} \approx 20$ km/s in CeIrIn$_5$ and $v_{Fb} \approx 4$ km/s in UPt$_3$ [22]. In Figure 3, the data normalized at $T_c$ is plotted as $\kappa / T$ vs $T$ for the two principal crystallographic directions: parallel and perpendicular to the high-symmetry $c$-axis (tetragonal in CeIrIn$_5$ and hexagonal in UPt$_3$). As can be seen, heat transport along the high-symmetry direction in both compounds is very similar: the temperature dependence of $c$-axis transport in CeIrIn$_5$ matches rather closely the transport in UPt$_3$. By contrast, the temperature dependence of heat transport in the basal plane is dramatically different in the two materials. In CeIrIn$_5$, $\kappa / T$ extrapolates to a large residual linear term, equal to 27 % of $\kappa / T$ at $T_c$. In UPt$_3$, a linear extrapolation of $\kappa / T$ data above 0.15 $T_c$, the lowest measured temperature in CeIrIn$_5$, yields a negative intercept at $T = 0$. UPt$_3$ data taken down to $T_c / 30$ extrapolates to a residual linear term no larger than $\kappa_{0b} / T \approx 0.2$ mW / K$^2$ cm, namely 1 % of $\kappa / T$ at $T_c$ [33]. Thirty times smaller than in CeIrIn$_5$, this is also one order of magnitude smaller than expected for a line node in this material. Indeed, applying to UPt$_3$ the theoretical estimate (Eq. 1) which works quantitatively very well for CeIrIn$_5$ [25], gives $\kappa_0 / T \approx 1.5$ mW / K$^2$ cm.

In other words, measured relative to the normal state, the anisotropy ratio $\kappa_c / \kappa_{ab}$ in the superconducting state of CeIrIn$_5$ grows rapidly with decreasing temperature, going from 1.0 at $T = 0.3\ T_c$ to 3.0 at $T = 0.15\ T_c$ and extrapolating to 10 or more as $T \to 0$, while it is more or less isotropic at all temperatures in UPt$_3$. In particular, at the lowest measured temperature of 16 mK, $\kappa_b \approx \kappa_c$ [33]. This is in striking contradiction with the presence of a line node in the basal plane. In the clean limit, at $T \approx T_c / 25$, the anisotropy should be enormous. To reconcile the small anisotropy observed in UPt$_3$ with the presence of a line node in the basal plane, a hybrid gap with *quadratic* point nodes along the *c*-axis was invoked, associated with the (1, *i*) state of the $E_{2u}$ representation in $D_{6h}$ symmetry [22]. However, such a gap would give universal conduction in both directions [4, 10], contrary to observation [32]. Indeed, measurements of heat transport in UPt$_3$ as a function of defect scattering (caused by electron irradiation), reveal that $\kappa_0 / T$ is not universal: in both directions, $\kappa_0 / T$ is found to grow with defect density (or residual resistivity). Again, for *J* || *c* this growth is remarkably similar to that observed in CeIrIn$_5$ for the same current direction [26]. In summary, the lack of universal conduction in UPt$_3$ for either current direction is strong evidence against a line node in the gap of this material. This leaves quite open the question of the gap structure in UPt$_3$.

**5. Summary**

We have outlined how heat transport at $T \to 0$ can shed light on the gap structure of a superconductor. The presence of a residual linear term is strong evidence for nodes, whose topology can be determined via the effect of impurity scattering. A line node yields universal heat conduction, independent of impurity scattering. Application of a magnetic field can reveal if an *s*-wave gap varies strongly around the Fermi surface, by exciting quasiparticles across a low minimum in the gap somewhere on the Fermi surface. We then showed how these principles elucidate the gap structure of three heavy-fermion superconductors: there is a line node in the basal plane of CeIrIn$_5$; there is no line node in the gap structure of UPt$_3$; part of the Fermi surface of CeCoIn$_5$ has a negligible gap. This kind of structure in the superconducting gap provides information on the *k*-dependence of the pairing mechanism.

**Acknowledgments**

We wish to thank our collaborators on studies of heat transport in superconductors over the years: Kamran Behnia, Benoit Lussier, Robert Gagnon, Christian Lupien, May Chiao, Brett Ellman, Etienne Boaknin, Cyril Proust, Rob Hill, Makariy Tanatar, Mike Sutherland, Dave Hawthorn, Johnpierre Paglione, Fil Ronning, Shiyan Li, and Nicolas Doiron-Leyraud. LT acknowledges the support of the Canadian Institute for Advanced Research and funding from NSERC, FQRNT, CFI and a Canada Research Chair.